\documentclass{article}
\usepackage{amsmath,amssymb,amsfonts,amsthm}
\usepackage{graphicx}

\newcommand{\be}{\begin{equation}}
\newcommand{\ee}{\end{equation}}
\newcommand{\beq}{\begin{eqnarray}} 
\newcommand{\eeq}{\end{eqnarray}}

\def\lsim{\hbox{ \raise.35ex\rlap{$<$}\lower.6ex\hbox{$\sim$}\ }}
\def\gsim{\hbox{ \raise.35ex\rlap{$>$}\lower.6ex\hbox{$\sim$}\ }} 

\begin{document}
\begin{flushleft}
KCL-PH-TH/2014-29
\end{flushleft}
\vskip.5truecm
\centerline{\bf Unweaving the Fabric of the Universe:}

\centerline{\bf The Interplay between Mathematics and Physics}

\vskip.5truecm
\centerline{M. Sakellariadou}

\vskip.5truecm
\centerline{\sl Department of Physics, King's College London, University of London,}

\centerline{\sl Strand WC2R 2LS, London, United Kingdom} 

\vskip.5truecm
\begin{abstract}

Our conventional understanding of space-time, as well as our notion of geometry, break down once we attempt to describe the very early stages of the evolution of our universe. The extreme physical conditions near the Big Bang necessitate an intimate interplay between physics and mathematics. The main challenge is the construction of a theory of quantum gravity, the long-sought unification of Einstein's general relativity with quantum mechanics. There are several attempts to formulate such a theory; they can be tested against experimental and observational results coming from high energy physics and astrophysics, leading to a remarkable interplay between gravity, particle physics and cosmology.
\end{abstract}

%\maketitle
%%%%%%%%%%%%%%%%%%%%%%
\section{\sl The plot} 
%%%%%%%%%%%%%%%%%%%
{\sl Unweaving the fabric of the universe}: a scientific question lying at the frontier between theoretical physics and mathematics, nourished by results from high energy physics and astrophysics.  

Einstein was attempting to unify gravity with sub-atomic forces, with the ultimate goal of weaving together all aspects of nature in a single fabric. His efforts were however unsuccessful, and his goal became the dream of all subsequent generations of theoretical physicists. At present, despite many efforts and the undoubtedly achieved progress, Einstein's dream, our dream, has still to be materialised. The inconsistency between General Relativity and Quantum Mechanics remains the main obstacle, yet to be resolved.

General Relativity, the well-known Albert Einstein's theory, describes warps in space-time, and accounts for the large-scale dynamics of the cosmos --- the dynamics of galaxies and clusters of galaxies, the dynamics of black holes and even the dynamics of our universe ---  with an extraordinary accuracy. General Relativity describes the way massive
objects curve space-time, turning a flat (dull) scenery into a curved (interesting) landscape of hills and basins. NASA's Gravity Probe B mission, launched in April 2004,  confirmed two fundamental predictions of the theory of General Relativity.  With the help of very precise gyroscopes, Gravity Probe B measured the warping of space-time around the Earth, as well as the amount of space-time the Earth pulls with it as it rotates. The former is known as the {\sl geodesic effect} and the latter as the {\sl lense-Thirring} or {\sl frame-dragging effect}. They were predicted by the theory of General Relativity to be equal to $ 6606 $ milliarcsec/year and $ 39.2 $ milliarcsec/year, respectively, while Gravity Probe B measurements lead to $ 6602 \pm  18 $ milliarcsec/year and $ 37.2 \pm 7.2 $ milliarcsec/year, respectively~\cite{Everitt:2011hp}. The accuracy is indeed remarkable.

The astonishing agreement between theoretical predictions based on General Relativity and measurements/observations disappears once we attempt to describe sub-atomic scales. Indeed,  once we zoom into the microcosm, the whole picture changes. At such tiny length scales, the behaviour and position of sub-atomic particles become uncertain; we penetrate the realm of Quantum Mechanics, full of uncertainties and probabilities. To realise Einstein's dream, our dream, we need to connect the curvature of space-time to the probabilities of events, relating General Relativity to Quantum Mechanics in a sole theory describing simultaneously the smooth macrocosm as well as the lumpy microcosm. This is the theory of Quantum Gravity.

Even though in most physical cases one can assume the validity of one of the two extreme cases, where either large length scales are relevant and hence General Relativity is the appropriate theory, or very small scales are instead the relevant ones and Quantum Mechanics plays the leading r\^ole, there are situations where one is forced to consider them both. The two well-known examples are the black hole's interior and the Big Bang phase.  I will briefly discuss them below.

For any star shinning in the dark sky, there is a series of nuclear reactions taking place in its interior. As time goes on however, it becomes more difficult for a star to meet the conditions needed for nuclear reactions to continue in its core.  Heavier elements are progressively formed in the core of a star, hence more extreme conditions are required for nuclear reactions to proceed. Eventually, the core of a star is made of heavy elements (e.g. iron) and nuclear reactions come to a stop.  Such an old star solves its energy crisis with a huge explosion scattering its outer layers in a structure known as supernova remnant. If the remaining dead
star is heavy enough, it collapses into a Black Hole, called {\sl black} because nothing, not even light, can escape from its event horizon. Even though General Relativity can perfectly well describe black hole dynamics, it cannot explain the physics of the black hole's interior: the theory blows up in the sense that  theoretical calculations based upon General Relativity lead to infinities. 

General Relativity describes very well the dynamics of our universe through Einstein's equations. Running these dynamic equations backwards in time, there is a finite time in the past when the universe must have been in a state of ultra-high curvature and mass density. This stage, called the Big Bang, characterised by a cosmological singularity which is unavoidable in classical General Relativity, took place about 13.8 billion years ago and is considered as the beginning of our observable universe. At that early time, all matter and energy of our present universe were confined into a tiny sub-atomic region of space. The classical theory of General Relativity cannot provide the mathematical tools and physical concepts that will allow us to describe the behaviour of gravity under such extreme, very high curvature and density, conditions. A quantum theory of gravity is indeed needed.  

To address questions, such as what is meant by the origin of the universe, what was the mechanism that led to its creation, whether its birth was an accidental process and whether it could have been avoided, or even what was before the Big Bang phase, requires some understanding about the state, structure and dynamics of space-time itself. Quantum Gravity is the theory aiming at explaining the (still unraveled) shape of space-time.
One could naively thought that space-time is structureless, but in reality it has instead an invisible structure, which we perceive as the force of gravity, and which determines our own trajectories, in the same way as waves guide a surfer on a rough sea, or bumps guide a downhill skier on a mountain slope.  
 
%%%%%%%%%%%%%%%%%%%%%%%%%%
\section{\sl The plot thickness}
%%%%%%%%%%%%%%%%%%%%%%%%%%%%%%%%%% 
Different approaches have been proposed so far as candidates for a theory of Quantum Gravity. They can be grouped into distinct classes categorised from either a particle physics motivation, or a general relativistic one, or even some geometric considerations. The main avenues searching for the theory of Quantum Gravity are:
\begin{itemize}
\item
String/M-theory:\\
A theory of all matter and forces, describing all particles as vibrating strings~\cite{Becker:2007zj}. 
\item
Loop Quantum Gravity:\\
Space is divided into (discrete) elements of volume. Loop Quantum Gravity represents the main alternative to String/M-theory~\cite{lqg}.
\item
Causal Dynamical Triangulations:\\
Space-time is considered as a collection of triangles~\cite{cdt}. This approach has emerged from  Euclidean Quantum Gravity, which views space-time as the result of a quantum average of all possible shapes, expressing the Feynman path integral as a sum over Riemannian metrics.~\cite{eqg}.
\item
Causal Sets:\\
Space-time is considered as discrete and events are related by a partial order representing causality relations between space-time events~\cite{causal-sets}.
\item
Noncommutative Geometry:\\
Geometry is considered as the product of a continuous 4-dimensional geometry for space-time and an internal 0-dimensional geometry for matter~\cite{ncg-book}. 
\end{itemize}
To test Quantum Gravity proposals, one can compare their cosmological predictions against the currently available astrophysical data. In this way, on the hand one can build a cosmological model inspired from some fundamental theory, instead of using an {\sl ad hoc} model, while on the other hand one can use the early universe as a very high energy laboratory  to test fundamental theories. 

To proceed, let us take a stroll through early universe cosmology. There is unquestionable evidence that our universe has been, and still is, expanding. Galaxies are running away from an observer, no matter where he/she is located, with a speed $v$ proportional to the distance $r$ between the galaxy and the observer. This is the well-known Hubble's law~\cite{hubble}:
\be
v=Hr~,
\ee
where $H$ denotes the rate of expansion of the universe and is called the Hubble parameter, with today's value $H_0=(67.3\pm 1.2) {\rm km}/{\rm s}/{\rm Mpc}$~\cite{Ade:2013zuv}, the Hubble constant. This law implies that a galaxy located at 3.26 million light years away, will be carried along the universal expansion of space at 73.8 km/sec. 

Our universe can be seen as the surface of a balloon whose radius increases with time. Assuming the {\sl cosmological principle} stating that the universe is homogeneous and isotropic on large scales, its geometry is adequately described by the standard cosmological model, the Friedmann-Lema\^itre-Robertson-Walker metric, which in comoving coordinates $r, \theta, \phi$ can be written as: 
\be
ds^2=-dt^2+a^2(t)\left[{dr^2\over 1-kr^2}+r^2d\theta^2+r^2\sin^2\theta d\phi^2   \right]~,
\ee
where $t$ stands for the cosmological time and $a(t)$ is the single dynamical quantity, called the scale factor, describing the size of the universe.
The parameter $k$ is the spatial curvature, with values $0, \pm 1$, determined by the overall energy density of the universe; its value will prescribe the fate of the universe.

The standard cosmological model, the {\sl Hot Big Bang Model},  provides a very successful description of the universe, but requires unnatural initial conditions, as for instance that the geometry of our universe is described by a flat euclidean space which is extremely homogeneous in all directions.  Given that the spatial curvature is a parameter, that can be interpreted as the difference between the average potential and kinetic energies of a spatial region, and that space-time is dynamical and  hence, according to General Relativity, it will curve in response of present matter sources, it is unnatural to expect that the universe is always extremely flat. 

In the Hot Big Bang model, gravity is described by General Relativity, hence the evolution of the universe is governed by Einstein's equations
\be
\label{einstein}
G_{\mu\nu}[g_{\mu\nu}]={8\pi G\over c^4}\sum_{i=1}^N T_{\mu\nu}^{(i)}~,
\ee
where $G_{\mu\nu}$ denotes the Einstein tensor, determined by the geometry specified by the metric tensor  $g_{\mu\nu}$, with greek letters running for the 3+1 space-time coordinates; $c, G$ stand for the speed of light and the Newtonian gravitational constant, respectively, while $T_{\mu\nu}^{(i)}$ denotes the energy-momentum tensor of a given source $i$. 
As current observational measurements have revealed the present universe is dominated by dark energy. The dark energy density parameter is $\Omega_\Lambda=0.686\pm 0.020$, while the matter density is just $\Omega_m =0.314\pm 0.020$. 
Einstein's equations relate the geometry of space-time (the left-hand-side), expressed through gravity, to the matter content (right-hand-side) through the energy momentum tensor within a particle physics content specified by the fluid component $i$. 

Einstein's equations, Eq.~(\ref{einstein}), reduce to ordinary nonlinear differential equations. The two independent equations for the 3 unknown quantities, the scale factor $a(t)$, the energy density $\rho(t)$ and the pressure $p(t)$, are
\be
H^2\equiv\left({\dot a\over a} \right)^2={8\pi G\over c^4}\sum_{i=1}^N \rho^{(i)}~,
\ee
and
\be
\dot\rho^{(i)}+3H\left(\rho^{(i)}+p^{(i)}\right)=0~,
\ee
where an over-dot stands for derivative with respect to cosmological time $t$.
The former is the Friedmann equation and the latter is the energy conservation or fluid equation. Hence to solve for the 3 unknown we need a third equation, which is 
the equation of state
\be
p^{(i)}=w\rho^{(i)}~,
\ee
relating the energy density $\rho^{(i)}$ of a fluid $i$ to its pressure $p^{(i)}$.

The rate of expansion of the universe will hence be determined by the type of
source that dominates its energy density. When the universe was dominated by relativistic particles, during the so called radiation-dominated era, the equation of state was $p=\rho/3$ leading to a power-law expansion $a(t)\propto t^{1/2}$.
Given that the energy density of relativistic particles behaves as $\rho\propto 1/a^4$, and that of non-relativistic (pressureless $p=0$) matter as $\rho\propto 1/a^3$, the latter will eventually dominate leading to the matter-dominated era during which the universe has the power-law $a(t)\propto t^{2/3}$ expansion.

To liberate the universe from the requirement of some of its specific initial conditions, the scenario of cosmological inflation has been proposed~\cite{inflation}. This scenario does not replace the Hot Big Bang model; it only enriches it during the early stages of our universe's infancy. According to the scenario of cosmological inflation, the universe in its very early stages was dominated by a fluid, that had a constant energy density and played the r\^ole of an effective cosmological constant (negative pressure with an equation of state $p=-\rho$) leading to an accelerated expansion $\ddot a>0$. This can be easily seen by writing down the acceleration equation
\be
{\ddot a\over a}=-{8\pi G\over c^4}(\rho +3p)~,
\ee
related to the Friedmann and fluid equations through Bianchi identities.
The period of inflation lasted from about $10^{-35}$ sec to about $10^{-32}$ sec after the Big Bang, during which the universe increased its size by approximately $10^{50}$. Note that today's energy density is about $10^{-47} {\rm GeV}^4$, inflation took place at an energy density of about $10^{64} {\rm GeV}^4$, while for comparison the energy density scale reached at the Large Hadron Collider at CERN is about $10^8 {\rm GeV}^4$.  Hence, by studying consequences of inflationary models we can learn about particle physics models in energy scales unreachable by any terrestrial colliders.

Cosmological inflation, this era of accelerated expansion, solves by construction the flatness and horizon problems, whilst it provides a mechanism for the origin of the seeds of initial matter perturbations leading to the observed large scale structure and the measured Cosmic Microwave Background temperature anisotropies measurements. 

However, despite the impressive agreement between inflationary predictions and astrophysical data, and the long years of research in this field, important open questions addressing the genericity of inflation~\cite{onset} and its relation to particle physics models (origin of the inflaton field)~\cite{inflaton} remain still unresolved. To address them we need Quantum Gravity, which will define the initial conditions at the beginning of inflation, whilst it is expected to modify the background equations considering Einstein's theory as an effective theory valid at lower energy scales. Moreover, it is hoped that Quantum Gravity may shed some light on the nature of the inflaton field. Note that there are unresolved puzzles of the Hot Big Bang model, referring to the initial singularity, baryon asymmetry and vacuum energy, which cannot be addressed by cosmological inflation. Quantum Gravity will at least evade the initial singularity issue. 

The ultimate test for any cosmological model is the comparison of its predictions with the Cosmic Microwave Background (CMB) temperature anisotropies data.
The early universe was extremely hot, so that there were no atoms, but only free electrons and nuclei. The free electrons acted as glue between photons and baryons through Thomson scattering. Hence at very early times, the universe was filled with a cosmological plasma of a tightly coupled photon-baryon fluid.
At the time of recombination, free electrons became bound into hydrogen and helium atoms, bringing the scattering of photons to an end.
At a redshift of about 1100 (hence within the matter-dominated era), about 380,000 years after the Big Bang, the universe  became transparent. Today, we observe a thin shell around us, the last scattering “surface”, where the overwhelming majority of photons last interacted with baryonic matter.

The Cosmic Microwave Background is just the afterglow radiation left over from the Hot Big Bang phase. Its extremely accurate black body spectrum, detected first time about 50 years ago by Penzias and Wilson, who heard some radio-signals coming from every direction in the sky and far beyond the Milky Way, provides the best confirmation of the validity of the Hot Big Bang cosmological model. The temperature of the Big Bang relic photons is extremely uniform with just tiny fluctuations. More precisely, the temperature of the universe has been progressively decreasing as a result of the universal expansion with current average temperature of $2.7255 \pm 0.0006$ K and tiny fluctuations of about $10^{-5}$. These tiny fluctuations have been observed by COBE satellite~\cite{cobe}  in 1992, the WMAP satellite~\cite{Komatsu:2010fb}  in 2003 and finally more recently by the PLANCK satellite~\footnote{http://www.esa.int/Planck}~\cite{Ade:2013zuv} in 2013 with an extreme accuracy.

Indeed, the observed universe is far from being homogeneous and isotropic on the scale of galaxies. The accepted mechanism is the amplification of initial perturbations by gravitational instability. The initial inhomogeneities can only grow during the matter-dominated era and their growth is slow and hence incompatible with the requirement of the recent (relatively speaking) structure formation under the observational constraint of tiny initial fluctuations revealed from the CMB temperature anisotropies measurements. Moreover, to reproduce the observational data, the two-point correlation function (called the {\sl power spectrum}~\footnote{The power spectrum gives the power of the variations of the fractional energy density of fluctuations as a function of a spatial scale.}) has to be close to scale invariance, a requirement that is just postulated within the Hot Big Bang model.
In the context of an inflationary scenario, quantum fluctuations of the gravitational field and the inflaton field can provide the source of the measured CMB temperature anisotropies and the observed large scale structure, leading to an almost scale invariant initial power spectrum.

%%%%%%%%%%%%%%%%%%%%%%%%%%%
\section{\sl The Plot Unfolds}
%%%%%%%%%%%%%%%%%%
Let me know briefly discuss how different approaches to Quantum Gravity can lead to cosmological models,  providing powerful tests of the validity of the Quantum Gravity proposals.
%%%%%%%%%%%%
\subsection{\sl String/M-theory}
%%%%%%%%%%%%%%%% 
String theory supposes that  matter consists of 1-dimensional objects, called strings. Different string vibrations (different string modes) would represent different particles, whilst splitting/joining of strings would correspond to different particle interactions. The strings can be either closed ones or open (without ends); the former represent gravitons, while the latter represent various matter fields. 
String theory has however a much richer spectrum that what has been originally thought. Assigning boundary conditions to the open strings, one can impose either Neumann or Dirichlet conditions. The former imply that the ends of the open strings move at the speed of light, whereas the latter ones mean that the ends of the open strings live on a surface, the D-brane, in space-time.
There is a well-known conjecture stating that all known string theories are different solutions to a more fundamental (11-dimensional) theory, called the M-theory. 

One may thus consider a higher dimensional bulk, within which branes of different dimensions can be embedded. Collisions and subsequent decays of higher dimensional branes can leave behind 3-dimensional branes (and anti-branes), one of which could play the r\^ole of our universe~\cite{3dim}, and 1-dimensional branes, called {\sl cosmic superstrings}.

Within the context of M-theory, brane cosmological models can be built, leading to the realisation of a brane inflation scenario. Since inflationary models based on Quantum Field Theory combined with General Relativity, can depend very sensitively on ultra-violet physics, inflation should indeed be studied within a ultra-violet complete theory, such as string theory.  Brane inflation models in string theory fall into two classes~\footnote{One should though keep in mind the caveats discussed in Ref.~\cite{rms}.}:
\begin{itemize}
\item D3/D7 inflation~\cite{Dasgupta:2002ew}, where  the attraction between
  the branes is due to the breaking of supersymmetry by the presence
  of a non-self-dual flux on the D7-brane.
\item Brane-antibrane inflation~\cite{dvalitye}, of
  which D3/$\overline {\rm D3}$~\cite{Kachru:2003sx} is the best
  studied example, where
  the attractive potential between the branes is due to
  warping.
\end{itemize}
Brane inflation models generically result in the production of cosmic superstrings. Their study is important since observational consequences of cosmic superstrings can serve as constraints on or signals of the underlying string theory model~\cite{cs-review}. These 1-dimensional branes play the r\^ole of their field theory analogues, cosmic strings~\cite{cosmic-strings}, which are 1-dimensional topological defects that may be formed during phase transitions followed by spontaneous symmetry breakdown of symmetries~\cite{Jeannerot:2003qv}, according to Kibble mechanism~\cite{cs-form}.  

Despite the various expected observational signatures of cosmic (super)strings, like 
\begin{itemize}
\item
gravitational lensing
\item 
gravitational waves (bursts or a stochastic background)
\item
cosmic microwave background temperature anisotropies
\item
pulsar timing
\end{itemize}
there is no observational evidence for their existence, while current data impose constantly stronger constraints.

%%%%%%%%%%%%%%%%%%%%%%%%%%%%%
\subsection{\sl Loop Quantum Gravity}
%%%%%%%%%%%%%%%%%
Loop Quantum Gravity postulates that space-time is itself quantised, composed of elementary discrete bits, called the {\sl quanta of space}, visualised as tiny 1-dimensional loops. These tiny quantised loops of gravitational fields weave an extremely fine fabric that is the 4-dimensional space-time of our physical world.  These loops are called {\sl spin networks} and once viewed over time they become what is called a {\sl spin foam}. 

Space has hence a discrete structure and cannot be indefinitely divisible into ever smaller (discrete) pieces.
Within Loop Quantum Gravity, a new formalism, the {\sl loop representation of Quantum General Relativity}, is used for the incorporation of quantum rules to the classical Einstein's theory of General Relativity.

Loop Quantum Gravity, belonging to the class of theories called {\sl Canonical Quantum Gravity}, is conceptually different than String/M-theory. It includes matter and forces but does not address the issue of unification of all forces and does not require extra spatial dimensions. 
Whereas in string theory the gravitational field can be seen as the sum of the background and the quantum field, in Loop Quantum Gravity there is no background, in the sense that 
loops are indeed the space. As a consequence, in Loop Quantum Gravity a state of space is described by a net of intersecting loops, and the theory is said to be {\sl background independent}. 

Freezing all but a finite degrees of freedom, one gets a mini-superspace model. Indeed, applying Loop Quantum Gravity in the context of homogeneous and isotropic backgrounds one gets a mini-superspace cosmological model, called Loop Quantum Cosmology~\cite{lqc}. Within Loop Quantum Cosmology, the cosmological model resulting from applying principles of the full Loop Quantum Gravity theory to cosmological settings, the quantum discreteness of the underlying geometry may
\begin{itemize}
\item
resolve the Big Bang singularity
\item
explain the temperature of Black Holes from first principles
\item
lead to new physics in the early universe.
\end{itemize}
Loop Quantum Cosmology can provide a framework to address the onset of inflation~[\cite{onset}c,e], and lead to modifications of the inflationary scenario with observable consequences, in particular regarding the superinflation phase and the characteristics of the CMB temperature perturbations~\cite{lqc-obs-infl}. 

%%%%%%%%%%%%%%%%%%%%%
\subsection{Noncommutative Geometry}
%%%%%%%%%%%%%%%%%%%%
To construct a quantum theory of gravity coupled to matter, one can either neglect matter altogether (as for instance in Loop Quantum Gravity), or consider instead that the interaction between gravity and matter is the most important aspect of the dynamics. The latter is indeed the philosophy followed in the context of Noncommutative Geometry, aiming at obtaining matter and gravity from (noncommutative) geometry~\cite{ncg}. The geometry is considered to be the tensor product of a continuous 4-dimensional geometry for space-time and an internal 0-dimensional geometry for the gauge content of the theory, namely the Standard Model of particle physics. The fruitful outcome of this approach is that considering gravity alone on the product space one obtains gravity and matter on the ordinary 4-dimensional space-time.

The conventional way to identify the geometry of a given object is by measuring distances between its points. There is however a different approach proposed more than a century ago by Weyl, who suggested to identify a membrane's shape~\footnote{The membrane is considered as a drum with the edges tied down to some solid material.} through the way it vibrates as one  bangs it. Studying the frequencies of the resulting fundamental tone and overtones, one may get information about the membrane's  geometry. Following Weyl's law, the largest frequencies of the sound of a membrane are basically determined by the area of the membrane and not by its shape, thus verifying Lorentz's conjecture.

In a paper under the intriguing title ``{\sl Can one hear the shape of a drum?}'', Kac has formulated the puzzle of whether one can reconstruct the geometry of a $n$-dimensional manifold (possibly with boundary) from the eigenvalues of the Laplacian on that manifold~\cite{kac}.  In other words, the relevant question is whether one (assuming he/she had perfect pitch) can deduce the precise shape of a drum just from hearing the fundamental tone and all overtones, even though one cannot really see the drum itself. Kac had stated that one most probably cannot hear the shape of a drum, nevertheless he investigated how much about the tambourine's shape one can infer from the knowledge of all eigenvalues of the wave equation that describes the vibrating object (the membrane). 

From the mathematical point of view, the question posed by Kac can be formulated as ``{\sl How well do we understand the wave equation?}''.
 For ordinary Riemannian geometry, the shape of the Laplacian does not fully determine the metric, so the shape of the drum cannot be heard, implying that we do not have a complete understanding of the (rather simple) wave equation.  But what about the product space considered within the Noncommutative Geometry context?

Every space vibrates at certain frequencies, hence one may consider the universe as a vibrating membrane, a tambourine. Let us also consider that space is identified with the product space of the 4-dimensional space-time and the internal space, defined by a mathematical object, called the {\sl spectral triple} given by the algebra of coordinates,
the Hilbert space and the Dirac operator corresponding to the
inverse of the Euclidean propagator of fermions.
The dynamics of the spectral triple are governed by a {\sl spectral action}~\cite{spectral-action} summing up all frequencies of vibration of the product space.
One may thus ask whether he/she can hear the shape of this product space. 
The spectrum of the Dirac operator together with the unitary equivalence class of its noncommutative spin geometry, fully determine the metric and its spin structure. Hence, one may indeed {\sl hear}  the shape of a spinorial drum.

Considering the spectral action and gravity coupled to matter, it was shown that one can obtain gravity and the Standard Model of elementary particles~\cite{Chamseddine:2006ep}. Thus, noncommutative spectral geometry offers a purely geometrical explanation for the Standard Model, the most successful particle physics model we still have at hand.

Let us hence consider General Relativity as an effective theory and  build a cosmological model based on Noncommutative Geometry along the lines of the spectral action. Given that this model lives by construction at very high energy scales, it provides a natural framework to construct early universe cosmological models. Studying these models one can test the validity of the Noncommutative Geometry proposal based on the spectral action, and in addition address some early universe open issues~\cite{ncsg-models}.

%%%%%%%%%%%%%%%%%%%%%%
\section{\sl Epilogue}
%%%%%%%%%%%%%%%
With no doubt, there are physical phenomena necessitating a theory of Quantum Gravity and the search for such a theory has become the {\sl obsession} of all theoretical physicists. Different proposals have been made and progress has been constantly achieved. I have discussed a few of the most developed approaches and argued that early universe cosmology can provide an appropriate context to test them.

 One may hence ask ``{\sl Among the various proposals, which is the right one to build a theory of Quantum Gravity?}'' and I feel that at present any of the proposals could be indeed the appropriate one, or any one could be just partially right, or even any one could be simply incorrect. Only nature can, and will, tell us! 

Even though we are not yet in a position to make any definite statement, we must admit that an enormous progress has been made and this will certainly help us to realise Einstein's dream, our own dream, and quantise gravity. 
Early universe cosmology can shed some light on the correct Quantum Gravity theory, unweaving the fabric of the universe and revealing the very early stages of its infancy.  

One may be tempted to ask ambitious questions like``{\sl What was before the Big Bang?}'' or ``{\sl How did the birth of our own universe take place?}'', etc. To be in a position to correctly address such questions we certainly need to make also further progress in mathematics. In other words, we may have to  formulate a new language to capture the structure of space-time and hence go beyond differential geometry, and we may have to construct a new set of variables in order to describe the constituents of space-time and those variables may not be given through the metric (as in General Relativity) or the connection (as in Loop Quantum Gravity). The dynamics of space-time will have to be expressed through a set of equations that are not the ones given within the context of General Relativity, which has hence to be considered as an effective theory. To achieve the target of unweaving the fabric of the universe, the intimate relation between mathematics and physics is stronger than ever.

Even though we cannot experience the Big Bang phase, and even less the pre-Big-Bang era, we can certainly try to construct the appropriate mathematical tools that will help us to unravel the mysteries and features of space-time. This is indeed the power of the intellectual mind and the beauty of theoretical physics.
%%%%%%%%%%%%%%%%%%%%%%%%%%%%%%

\end{document}